\documentclass{PoS}
\usepackage{bbm}
\usepackage{amssymb}
\usepackage{ifpdf}
\usepackage{color}
\usepackage{graphicx,epsfig}
\newcommand{\1}{{\mathbbm{1}}}

\title{Real-time simulation of dissipation-driven quantum systems}

\ShortTitle{Real-time simulations}

\author{\speaker{Debasish Banerjee}\\
        NIC, DESY, Platanenallee 6, 15738 Zeuthen, Germany\\
        E-mail: \email{debasish.banerjee@desy.de}}

\author{Florian Hebenstreit, Uwe-Jens Wiese 
      \thanks{The  research  leading  to  these  results
     has received funding from the Schweizerischer Nationalfonds
     and  from  the  European  Research  Council  under 
     the  European  Union's  Seventh  Framework  Programme
     (FP7/2007-2013)/ ERC grant agreement 339220.}\\
        Albert Einstein Center, Institute for Theoretical Physics,
        Bern University, 3012 Bern, Switzerland\\
        E-mail: \email{hebenstreit@itp.unibe.ch, wiese@itp.unibe.ch}}

\author{Fu-Jiun Jiang\\
        Department of Physics, National Taiwan Normal University,\\
        88, Sec. 4, Ting-Chou Rd., Taipei 116, Taiwan\\
        E-mail: \email{fjjiang@ntnu.edu.tw}}

\author{Mark Kon\\
        Department of Mathematics, Boston University, 
        Boston, Massachusetts, USA\\
        E-mail: \email{MKON@bu.edu}}

\abstract{We set up a real-time path integral to study the evolution
         of quantum systems driven in real-time completely by the
         coupling of the system to the environment. For specifically chosen
         interactions, this can be
         interpreted as measurements being performed on the system. 
         For a spin-1/2 system, in particular, when the measurement 
         results are averaged over, the resulting sign problem completely
         disappears, and the system can be simulated with an efficient 
         cluster algorithm.}

\FullConference{The 33rd International Symposium on Lattice Field Theory\\
		14 -18 July 2015\\
		Kobe International Conference Center, Kobe, Japan*}

\begin{document}

\section{Introduction}

\vspace{-0.3cm}
  The generally established way to study the static properties of 
large quantum systems is using Monte Carlo methods. However, the notorious
'sign problem' may arise in cases
where the probability sampling over the various stochastically generated
configurations have negative (or complex) weights.
The particularly exciting branch of theoretical
physics dealing with
the real-time evolution of large quantum systems, gets excluded from the
reach of Monte Carlo methods since the configurations contributing to
the real-time path integral have complex weights. Moreover, 
diagonalizing the Hamiltonian is impossible in practice, due to the
exponential growth of the Hilbert space with the system size. 
Matrix product states provide a basis
for simulating the real-time evolution of quantum systems, 
based on the density matrix renormalization group \cite{Whi92}. 
They are successful only for gapped 1-D systems for moderate time-intervals
\cite{Vid03,Whi04,Ver04,Zwo04,Dal04,Bar09,Piz13}. 
Sign problems also occur in Euclidean time simulations of Quantum
Chromodynamics at non-zero baryon chemical potential and the 
fermionic Hubbard model away from half-filling. Frustrated
systems and actions with non-zero topological terms
again face the same problem. 
The different physical origins of the different sign problems suggest that
there might not be a unique solution applicable to all the
cases \cite{Tro05}.

 One reason why classical computers have problems to
simulate quantum systems - especially in real-time - is because
quantum entanglement is not easily representable, let
alone computable, as classical information in conventional 
computers. This already led Feynman to propose the use of specially
designed quantum devices to mimic quantum systems \cite{Fey82}, 
very much along 
the lines of the usage of toy aeroplane models to study actual problems
of aircraft manoeuvring in real-time flight situations. This
so-called analog computing differs from a digital
computer, which is a machine capable of computing an
answer to a problem according to an algorithm.
Ever since the experimental realization of Bose-Einstein 
condensation \cite{And95,Dav95}, quantum optics has progressed 
by leaps and bounds, 
and the degree of control available for ultracold atomic
systems is truly remarkable. The bosonic Hubbard model
has been implemented with well-controlled ultracold atoms in an 
optical lattice \cite{Gre02}, and several aspects of this quantum 
simulation have been verified by comparison with accurate
quantum Monte Carlo simulations \cite{Tro10}. Digital \cite{Llo96} and
analog \cite{Jak98} quantum simulators are widely discussed in
atomic and condensed matter physics 
\cite{Cir12,Lew12,Blo12,Bla12,Asp12,Hou12}, and more
recently also in a particle physics context 
\cite{Kap11,Szi11,Zoh12,Ban12,Ban13,Zoh13,Tag13a,Tag13b,Wie13}.

 However, quantum simulators are not always universally applicable, and
even more importantly they are not yet precision instruments. Therefore, 
the study of real-time evolution of quantum
systems using classical computers remains an open important challenge.
Closed quantum systems tend to evolve 
into complicated entangled states under the action of the Hamiltonian
 in real time rendering their simulation difficult. 
A different approach is to consider open quantum systems,
as they undergo decoherence due to their continuous 
interaction with the environment.

In this work, we have developed
a method to simulate the dynamics of large quantum systems, driven 
entirely by the measurement process of the
total spin $(\vec S_x + \vec S_y)^2$ of pairs of 
spins $\frac{1}{2}$ at nearest neighbors $x$ and $y$. These measurements
can also be interpreted as a dissipative coupling to the 
environment, especially when this interaction takes place 
stochastically. As we shall see, the system is driven from a given
equilibrium initial state to a new equilibrium final state for late real-times. 
This is the first example where the real-time evolution of a
large strongly coupled quantum system can be studied over
arbitrarily large periods of real time in any spatial dimension. 
%

\section{Formulation of the problem}

\vspace{-0.3cm}
 To begin with, we outline the path integral representation for the
real-time process driven by measurements. Consider a general quantum
system with a Hamiltonian, whose evolution in real time from $t_k$ to $t_{k+1}$
is described by the operator $U(t_{k+1},t_k) = U(t_k,t_{k+1})^\dagger$.
An observable $O_k$ with an eigenvalue $o_k$ is measured at time $t_k$
($k \in \{1,2,\dots,N\}$). The measurement projects
the state of the system to the subspace of the Hilbert space spanned by the 
eigenvectors of $O_k$ with eigenvalue $o_k$, and is represented by 
the Hermitean operator $P_{o_k}$. Starting from an initial 
density matrix $\rho_0 = \sum_i p_i |i\rangle\langle i|$ (with 
$0 \leq p_i \leq 1$, $\sum_i p_i = 1$) at time $t_0$, the probability to reach a
final state $|f\rangle$ at time $t_f$, after a sequence of $N$ measurements with
results $o_k$, is then given by \cite{Gri84}

\vspace{-0.6cm}
\begin{eqnarray}
\label{probrhof}
&&p_{\rho_0 f}(o_1,o_2,\dots,o_N) = \\ \nonumber
&&\sum_i \langle i|U(t_0,t_1) P_{o_1} U(t_1,t_2) P_{o_2} \dots P_{o_N} U(t_N,t_f)
|f\rangle \langle f|U(t_f,t_N) P_{o_N} \dots P_{o_2} U(t_2,t_1) P_{o_1} U(t_1,t_0)|i\rangle p_i.
\end{eqnarray}
In general, matrix elements for both the time-evolution and the projection operators 
are complex, and cannot be simulated by Monte Carlo methods. 
However, classical measurements should disentangle the
quantum system and reduce the sign problem. For simplicity,
we consider systems completely driven by measurements, i. e.  $U(t_k,t_{k+1}) = \1$. 
To construct the path integral, we insert 
$\sum_{n_k}|n_k\rangle\langle n_k| = \1$ into the first 
factor and independently $\sum_{n_k'}|n_k'\rangle\langle n_k'| = \1$ into the 
second factor in eq.\ (\ref{probrhof}), at all times $t_k$. After some
rearranging, we arrive at the following expression for the
real-time path integral along the Keldysh contour leading from $t_0$ to $t_f$ 
and back \cite{Sch61,Kel65}: 

\vspace{-0.6cm}
\begin{eqnarray}
&& p_{\rho_0 f}(o_1,o_2,\dots,o_N) = 
 \sum_i p_i \sum_{n_1,n_1'} \dots \sum_{n_N,n_N'} 
\prod_{k=0}^N \langle n_k n_k'|P_{o_k} \otimes P_{o_k}^*|n_{k+1} n_{k+1}'\rangle.
\end{eqnarray}
Here
$\langle n_k n_k'|P_{o_k} \otimes P_{o_k}^*|n_{k+1} n_{k+1}'\rangle =
\langle n_k|P_{o_k}|n_{k+1} \rangle \langle n_k'|P_{o_k}|n_{k+1}'\rangle^*$,
$\langle n_0 n_0'| = \langle i i|$, and $|n_{N+1} n_{N+1}'\rangle = |f f\rangle$. 
Further, if the intermediate measurements are of no interest (as is usually the case
for dissipation), we can sum over them to arrive at the following result:

\vspace{-0.6cm}
\begin{equation}
 p_{\rho_0 f} = \sum_{o_1} \sum_{o_2} \dots \sum_{o_N} p_{\rho_0 f}(o_1,o_2,\dots,o_N) 
 = \sum_i p_i \sum_{n_1,n_1'} \dots \sum_{n_N,n_N'} 
\prod_{k=0}^N \langle n_k n_k'|\widetilde P_k|n_{k+1} n_{k+1}'\rangle,
\label{pi}
\end{equation}
A dissipative system continuously interacting with an environment can be described by
the Lindblad equation. For the system under consideration, the Lindblad operators \cite{Kos72,Lin76} 
$L_{o_k} = \sqrt{\varepsilon \gamma} P_{o_k}$, obey
$(1 - \varepsilon \gamma N) \1 + \sum_{k,o_k} L_{o_k}^\dagger L_{o_k} = \1$, where
$\gamma$ determines the probability of measurements per unit time. The index $k$ now
labels the Lindblad operators at any fixed instant of time $t_k$. The Lindblad
equation is:

\vspace{-0.4cm}
\begin{equation}
\partial_t \rho = \frac{1}{\varepsilon} \sum_{k,o_k} \left(
L_{o_k} \rho L_{o_k}^\dagger - \frac{1}{2} L_{o_k}^\dagger L_{o_k} \rho - 
\frac{1}{2} \rho  L_{o_k}^\dagger L_{o_k} \right) 
=\gamma \sum_k (\sum_{o_k} P_{o_k} \rho P_{o_k} - \rho).
\end{equation}
The second equation is obtained in the continuous time limit, $\epsilon \to 0$.
Based on the Lindblad equation, we can again arrive at a path integral
expression similar to the one in eqn. (\ref{pi}).
 
For illustration, lets consider two spins $\frac{1}{2}$, $\vec S_x$
and $\vec S_y$. The total spin $S$ eigenstates $|S S^3\rangle$ (with
3-component $S^3$) are: $|1 1\rangle = | \uparrow\uparrow \rangle$,
$|1 0\rangle = \frac{1}{\sqrt{2}}( | \uparrow\downarrow \rangle + | \downarrow\uparrow \rangle)$,
$|1 -1\rangle = | \downarrow\downarrow \rangle $, and
$|0 0\rangle = \frac{1}{\sqrt{2}}(| \uparrow\downarrow \rangle - | \downarrow\uparrow \rangle)$.
The projection operators corresponding to a measurement 1 or 0 of the total spin
are then given by $P_1 = |1 1\rangle \langle 1 1| + |1 0\rangle \langle 1 0| +
|1 -1\rangle \langle 1 -1|$ and $P_0 = |0 0\rangle \langle 0 0|$, such that

\begin{equation}
P_1 = \left(\begin{array}{cccc} 
1 & 0 & 0 & 0 \\ 0 & \frac{1}{2} & \frac{1}{2} & 0 \\
0 & \frac{1}{2} & \frac{1}{2} & 0 \\ 0 & 0 & 0 & 1 \end{array}\right), \quad
P_0 = \left(\begin{array}{cccc} 
0 & 0 & 0 & 0 \\ 0 & \frac{1}{2} & - \frac{1}{2} & 0 \\
0 & - \frac{1}{2} & \frac{1}{2} & 0 \\ 0 & 0 & 0 & 0 \end{array}\right).
\end{equation}
Note that the negative entries in $P_0$ would give rise to the sign
problem in a Monte Carlo simulation. However, when the measurement
results are not distinguished,
 $\widetilde P = P_1 \otimes P_1^* + P_0 \otimes P_0^*$
all the matrix elements of the operator are non-negative. 
Furthermore, a very efficient loop-cluster algorithm
has been designed to simulate the system \cite{Ban14}.

\section{Simulating a large quantum system in real-time}

 \vspace{-0.3cm}
 The above example can be straightforwardly generalized 
to the case of a large quantum spin system in any dimension. We have
implemented this process for a spin $S=\frac{1}{2}$ system on a square lattice
of size $L \times L$ with periodic boundary conditions. The measurement 
process involves the nearest-neighbor spins, 
and is implemented in four steps. The first step measures
all the neighboring spins in the 1-direction, at $x = (x_1,x_2)$ 
and $y = (x_1+1,x_2)$ with even $x_1$, simultaneously. The second step looks
at the spins at $(x_1,x_2)$ and $(x_1,x_2+1)$ with even $x_2$. 
In a third and fourth measurement step, 
the total spins of pairs with odd $x_1$ and $x_2$ are measured. This 
is repeated an arbitrary number of times $M$, such
that the total number of measurements is $N = 4M$. As a Lindblad
process, the exact ordering of the measurement is irrelevant, and 
the measurements are stochastic. 
We use the Hamiltonian of the Heisenberg anti-ferromagnet 
$H = J \sum_{\langle xy \rangle} \vec S_x \cdot \vec S_y$,
to prepare an initial density matrix $\rho_0 = \exp(- \beta H)$.
A cluster algorithm is used to update the whole system, including
both the Euclidean and real-time branches as explained in \cite{Ban14}.

\begin{figure}[tbp]
\begin{center}
\includegraphics[width=0.4\textwidth]{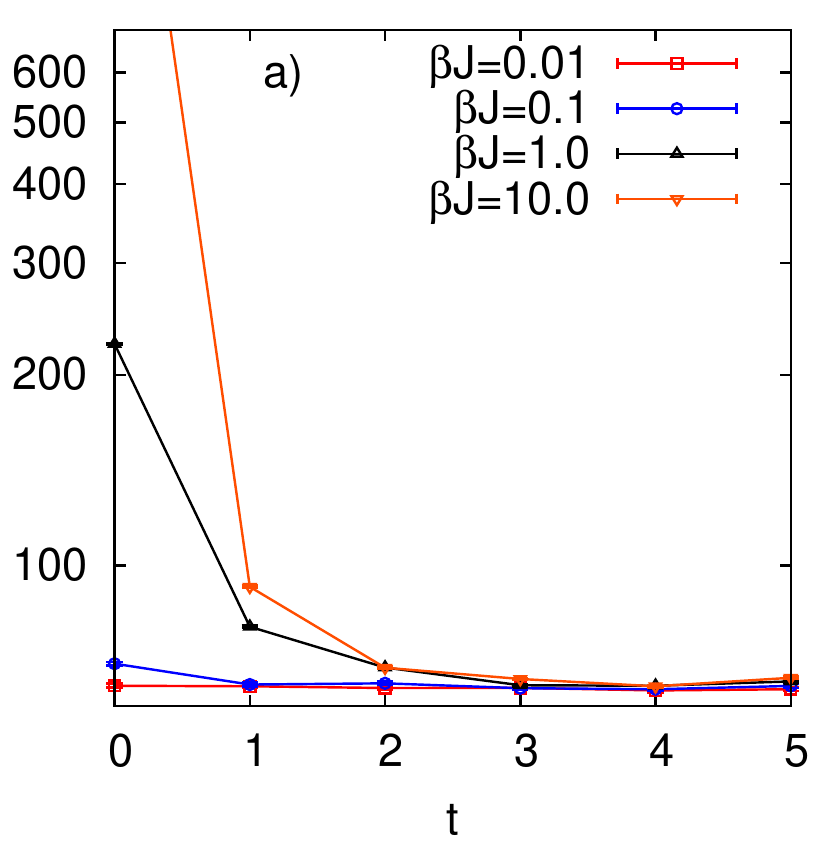}
\includegraphics[width=0.45\textwidth]{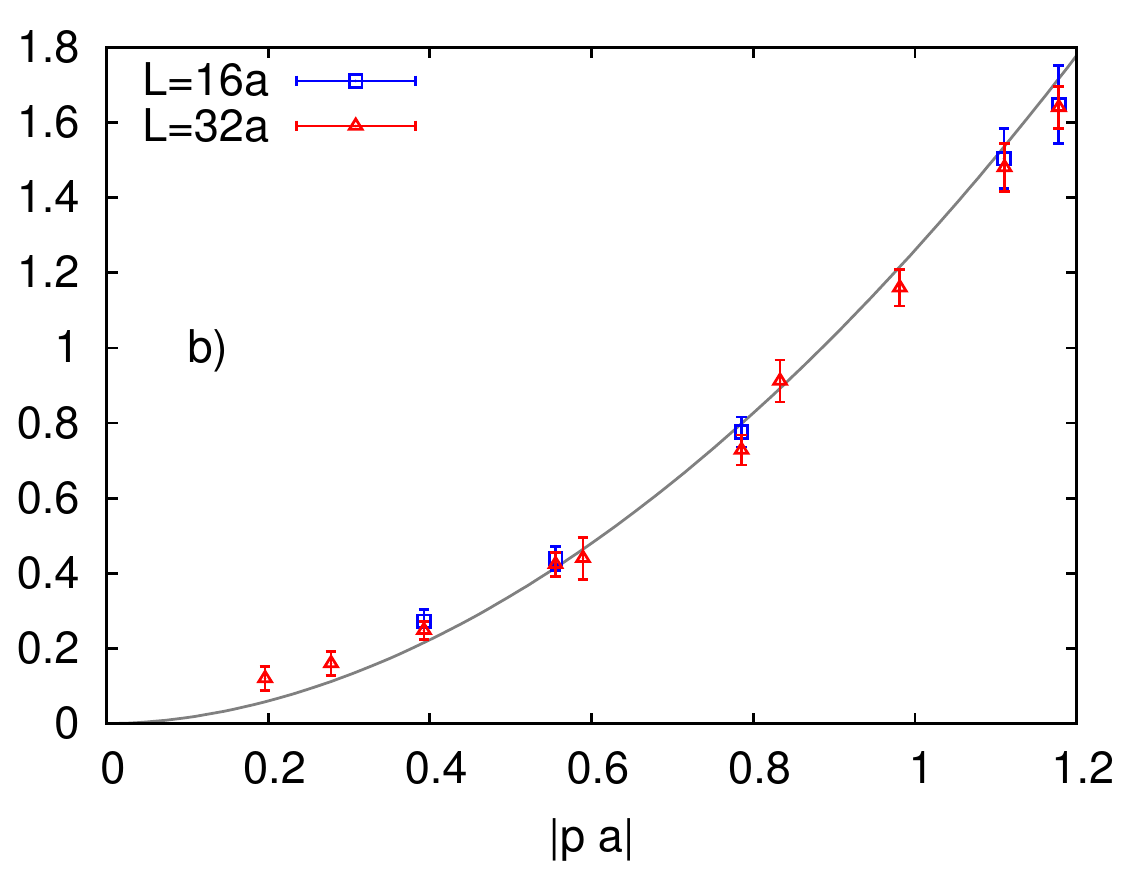} \\
\includegraphics[width=0.6\textwidth]{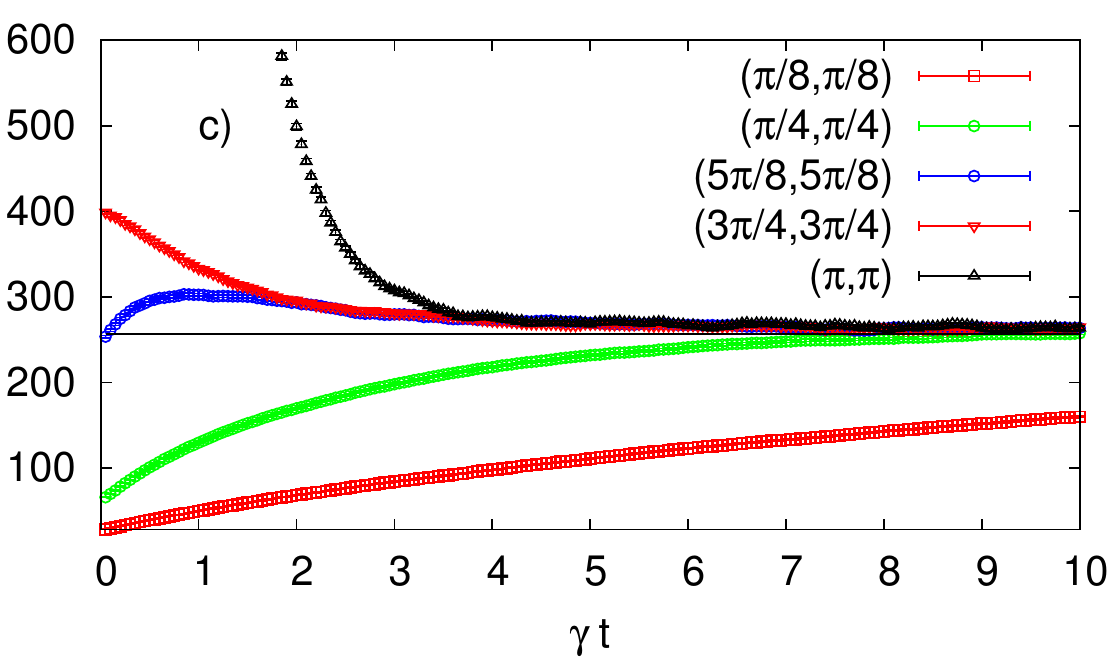}
\end{center}
\caption{\textit{a) Real-time evolution of 
$\langle M_s^2 \rangle$ driven by discrete measurements, for 
$\beta J = 0.01$, $0.1$, $1$, and $10$, for $L = 16 a$. b) Inverse equilibration
time $1/[\gamma \tau(p)]$ as a function of $|p|$ for $L = 16 a$, $\beta J = 40$,
and $L = 32 a$, $\beta J = 80$. c) Evolution of the Fourier modes 
$\langle |\widetilde S(p)|^2 \rangle$ for the Heisenberg anti-ferromagnet driven 
by a continuous Lindblad process. The initial ensemble is at a low temperature 
$\beta J = 5 L/2 a$, where $a$ is the lattice spacing.}}
\end{figure}

 At $T=0$, the global $SU(2)$ symmetry of the Heisenberg model
is spontaneously broken, and this gives a large signal in the staggered 
magnetization $M_s = \sum_x (-1)^{x_1+x_2} S^3_x$ of the initial state. 
Discrete measurements quickly destroy this order, and drive the system to
a new steady state as shown in Fig.\ 1a. A Lindblad process
also has the same effect, but since all the spin pairs 
are not affected simultaneously, the approach to the new steady state
happens more slowly, and is dependent on the dissipative coupling
strength. The Fourier modes, 
$\widetilde S(p) = \sum_x S^3_x \exp(i p_1 x_1 + i p_2 x_2)$, $p = (p_1,p_2)$,
show a similar behavior (Fig.\ 1c). The symmetry
breaking is signalled by a large condensate signal at the Fourier mode $(\pi,\pi)$, 
and in the nearby modes around this one.
 The uniform magnetization, $\vec M = \sum_x \vec S_x$, on the other hand,
is conserved by both the 
Hamiltonian and the measurement process. Therefore, while 
the mode $(0,0)$ does not equilibrate at all, the nearby low-momentum
modes approach the final state very slowly, according to
$\langle |\widetilde S(p)|^2 \rangle \rightarrow A(p) + B(p) \exp(- t/\tau(p))$. 
For small momenta, the equilibration 
process is diffusive: 
$1/[\gamma \tau(p)] = C |p a|^r$, $C = 1.26(8)$, $r = 1.9(2)$ (Fig.\ 1b).

 The dissipation process conserves not 
only the total spin, but also the lattice translation and rotation
symmetries. The final density matrix to which the system is driven, is
therefore constrained by these symmetries, and is proportional to 
the unit matrix in each of these sectors. This can be analytically
computed to give $A(p) = L^4/(L^2-1)$, shown as the horizontal line
in Fig.\ 1c. This is a stable fixed point of the full dynamics (Hamiltonian
and the Lindbladian) at $T=\infty$, and acts as a universal attractor 
for a large class of dissipative process \cite{Flo15}.

 Another interesting issue is whether a phase transition occurs at a finite
instant in real time. While the initial state breaks $SU(2)$ symmetry spontaneously,
the final state has a volume independent $\langle M_s^2 \rangle/L^2$, 
indicating a restored $SU(2)$ spin symmetry. This implies that
the system passes through a (second-order) phase transition. However,
since the Lindblad process takes the system out of thermal equilibrium, 
this phase transition is not expected to fall into any of the standard
dynamic universality classes \cite{Hoh77}. To study this, we plot 
$\langle M_s^2 \rangle/L^4$ as well as the Binder ratio
$\langle M_s^4 \rangle/\langle M_s^2 \rangle^2$ for $\beta J = 2L/3a$
in Fig.\ 2a and 2b. The finite volume curves do not intersect, but their
point of inflection moves to later and later real times with
increasing spatial volume. The staggered magnetization density
${\cal M}_s$, and the length scale $\xi = c/(2 \pi \rho_s)$
(where $c$ is the spin-wave velocity and $\rho_s$ is 
the spin stiffness) are obtained as a function of time, by a fit to
the equation
\begin{equation}
\langle M_s(t)^2 \rangle = \frac{{\cal M}_s(t)^2 L^4}{3} 
\sum_{n=0}^3 c_n \left(\frac{\xi(t)}{L}\right)^n,
\end{equation}
which implicitly defines ${\cal M}_s(t)$ and $\xi(t)$.The constants
$c_0 = 1$, $c_1 = 5.7503(6)$, $c_2 = 16.31(2)$, $c_3 = - 84.8(2)$,  
accurately determined at $t = 0$, are assumed to be time-independent
since they are related to the spatial geometry of the system. The
exponential decay of the order parameter 
${\cal M}_s(t) = {\cal M}_s(0) \exp(-t/\tau)$, 
(with ${\cal M}_s(0) = 0.30743(1)/a^2$ \cite{San08,Ger09})
suggests that it takes an infinite amount of time for the
order to disappear completely and hence for the
phase transition to be completed. 

\begin{figure}[tbp]
\begin{center}
\includegraphics[width=0.4\textwidth]{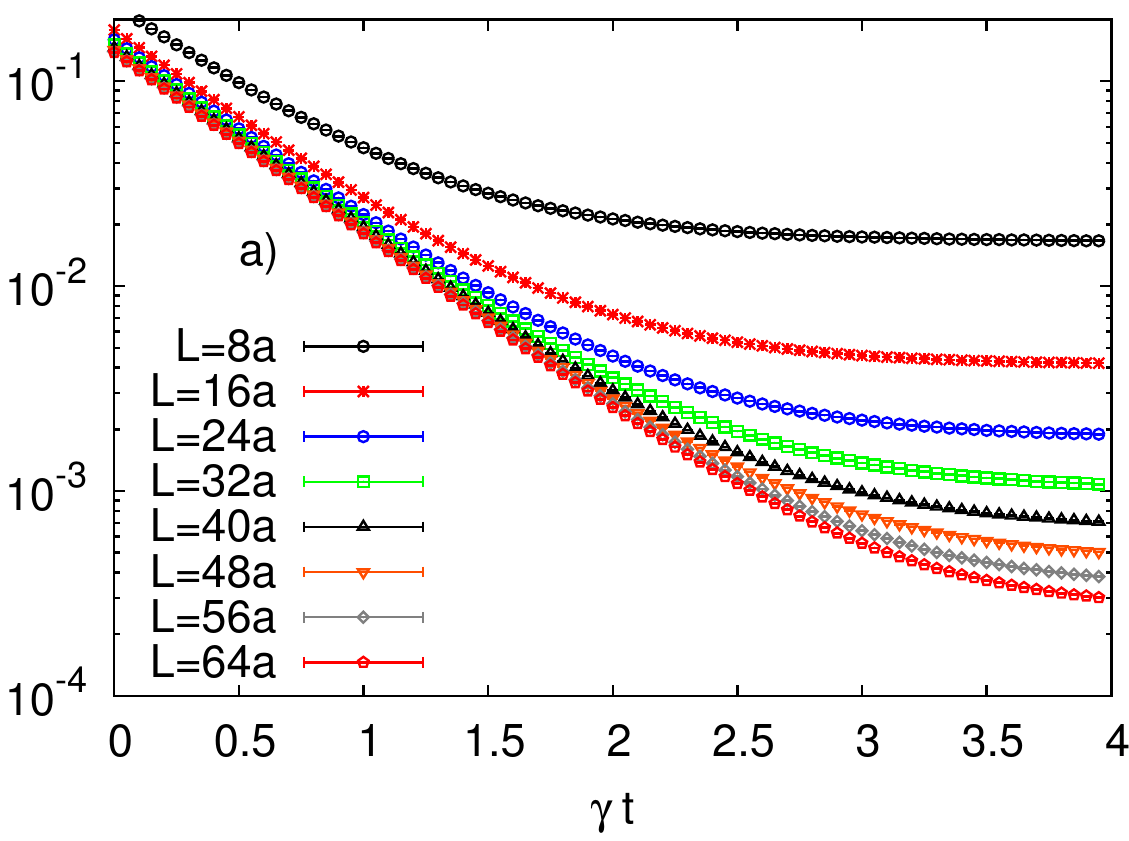}
\includegraphics[width=0.4\textwidth]{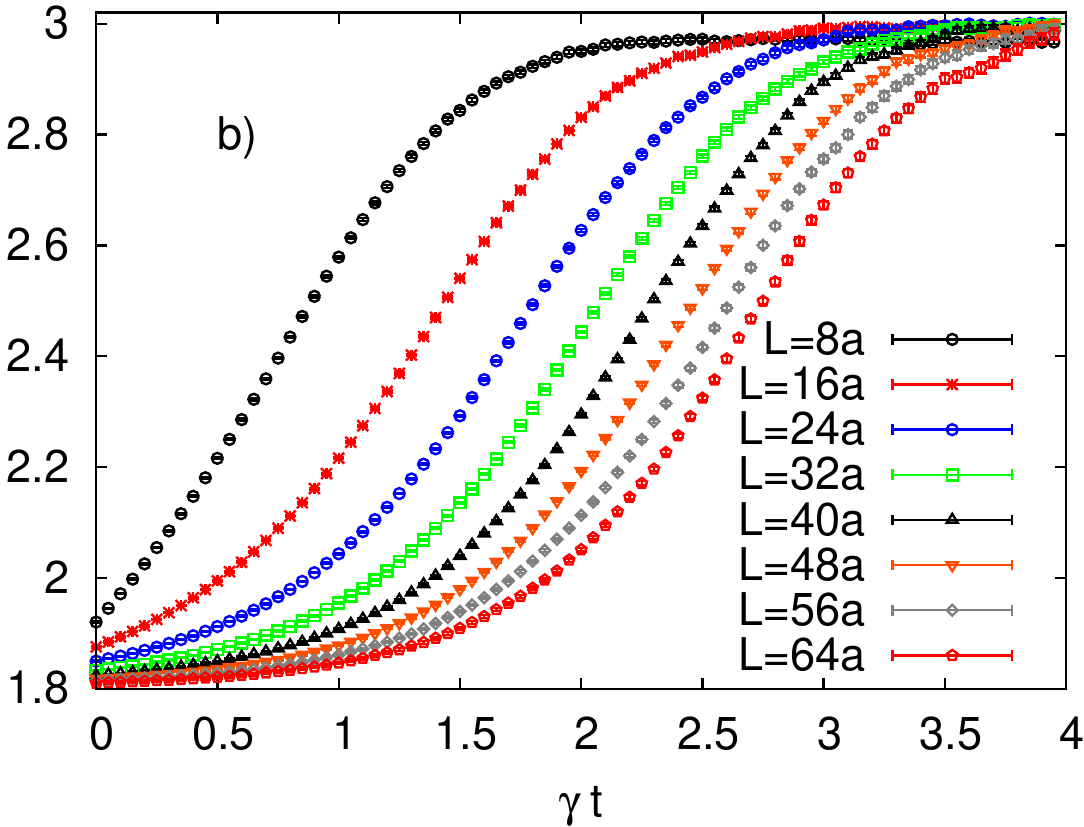} \\
\includegraphics[width=0.4\textwidth]{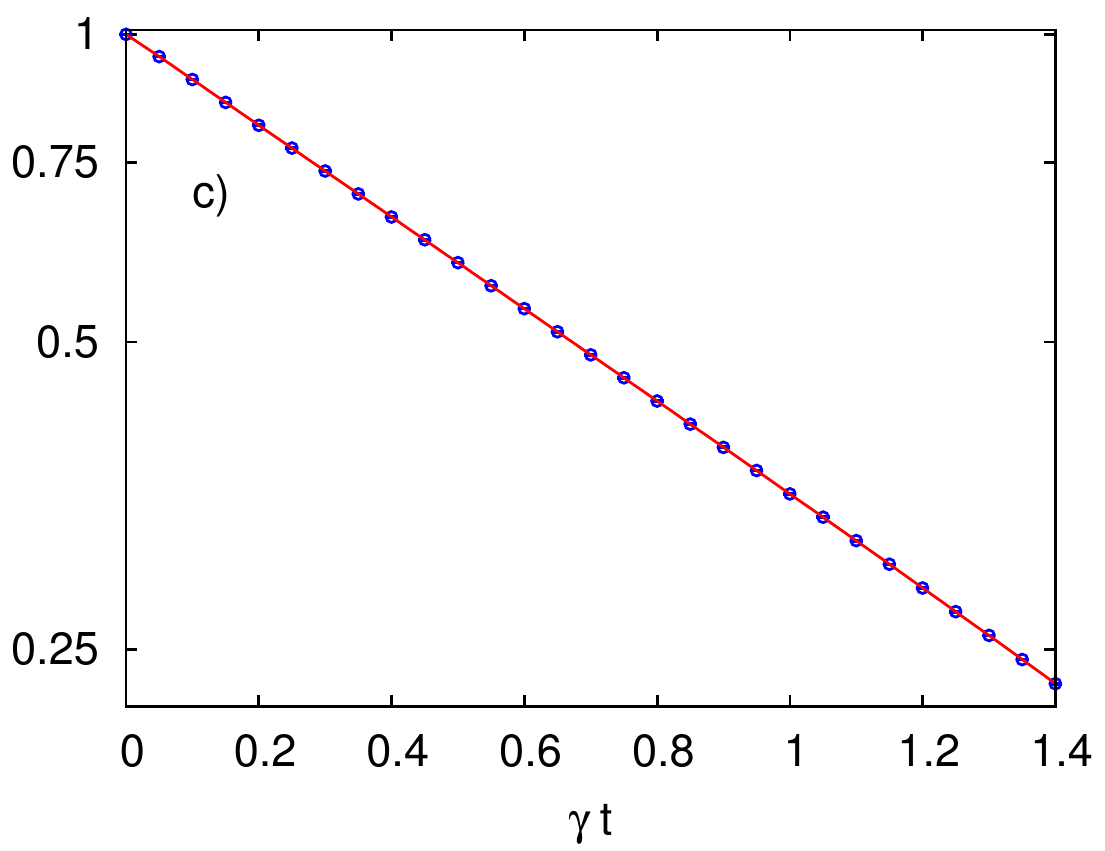}
\includegraphics[width=0.4\textwidth]{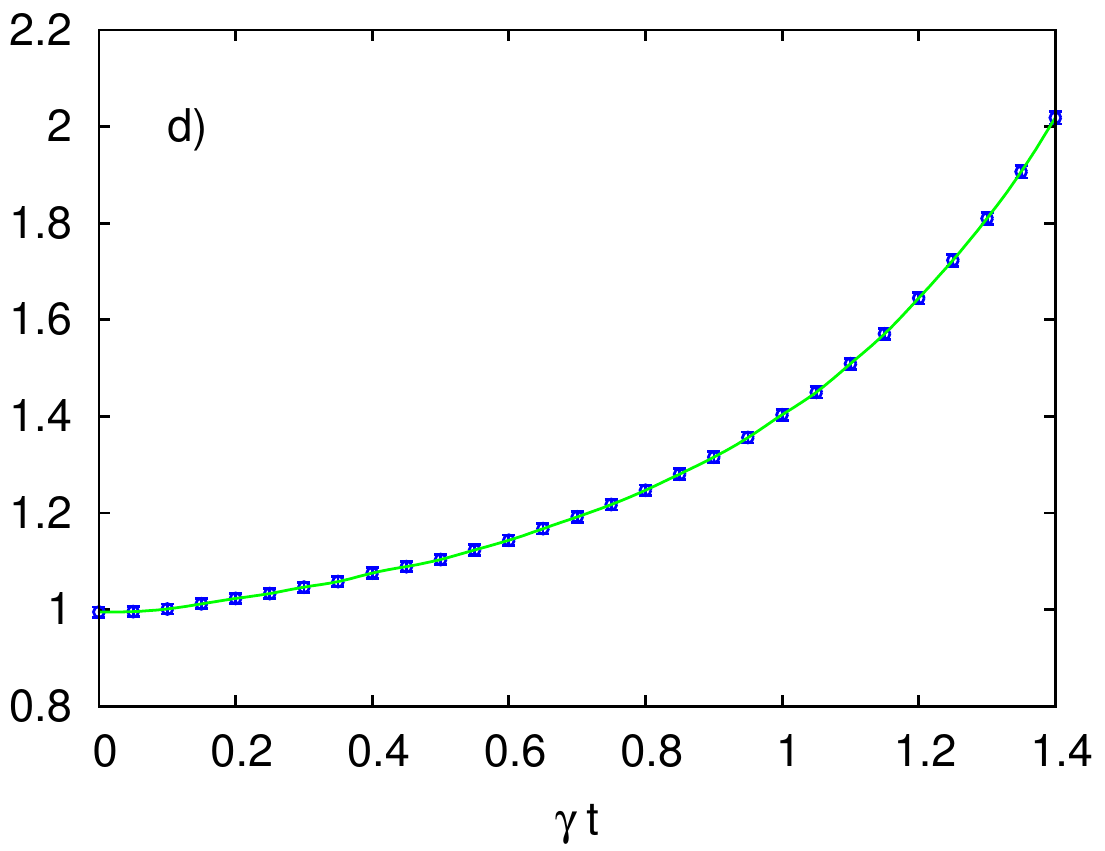}
\end{center}
\caption{[Color online] \textit{a) $\langle M_s^2 \rangle/L^4$ and b) Binder
ratio $\langle M_s^4 \rangle/\langle M_s^2 \rangle^2$ as functions of time for
$L/a = 12,\dots,48$, $\beta J = 8,\dots,30$. Evolution of c) 
${\cal M}_s(t)/{\cal M}_s(0)$ and d) $\xi(t)/\xi(0)$.}}
\end{figure}

\section{Conclusion and Outlook}
 
 \vspace{-0.3cm}
 We have demonstrated that for certain measurement or dissipative
processes, we can simulate the real-time evolution of a large
quantum system on classical computers. The processes under
consideration drive initial states (or density
matrices) into a new state with only short-range correlations.
However, the phase transition into a final disordered steady state without
spontaneous symmetry breaking is completed only in an infinite amount of time.  

 Extending these investigations, we have also studied several other
measurement processes and initial states given by the ground states of
different models \cite{Flo15}. 
All these cases show a diffusive behavior
of a conserved quantity (i.e., the staggered or the uniform magnetization).
Further, the non-equilibrium
transport of magnetization in large open systems has also been studied
with our method \cite{Flo15a}. 
A study of competing Lindblad 
processes, and the inclusion of a part of the Hamiltonian along with the
dissipation, as well as extension of this formalism to fermionic systems.
are currently under investigation.


\begin{thebibliography}{99}

\bibitem{Whi92}
S.\ R.\ White, Phys.\ Rev.\ Lett.\ 68 (1992) 2863.

\bibitem{Sch05}
U.\ Schollw\"ock, Rev.\ Mod.\ Phys.\ 77 (2005) 259.

\bibitem{Vid03}
G.\ Vidal, Phys.\ Rev.\ Lett.\ 91 (2003) 147902.

\bibitem{Whi04}
S.\ R.\ White and A.\ E.\ Feiguin, Phys.\ Rev.\ Lett.\ 93 (2004) 076401.

\bibitem{Ver04}
F.\ Verstraete, I.\ I.\ Garcia-Ripoll, and J.\ I.\ Cirac, 
Phys.\ Rev.\ Lett.\ 93 (2004) 207204.

\bibitem{Zwo04}
M.\ Zwolak and G.\ Vidal, Phys.\ Rev.\ Lett.\ 93 (2004) 207205.

\bibitem{Dal04}
A.\ J.\ Daley, C.\ Kollath, U.\ Schollw\"ock, and G.\ Vidal, 
J.\ Stat.\ Mech.: Theor.\ Exp.\ P04005 (2004).

\bibitem{Bar09}
T.\ Barthel, U.\ Schollw\"ock, and S.\ R.\ White, Phys.\ Rev.\ B79 (2009)
245101.

\bibitem{Piz13}
I.\ Pizorn, V.\ Eisler, S.\ Andergassen, and M.\ Troyer, New J.\ Phys.\ 16 (2014) 073007.

\bibitem{Tro05}
M.\ Troyer and U.-J.\ Wiese, Phys.\ Rev.\ Lett.\ 94 (2005) 170201.

\bibitem{Fey82}
R.\ P.\ Feynman, Int.\ J.\ Theor.\ Phys.\ 21 (1982) 467.

\bibitem{And95}
M.\ H.\ Anderson  et. al., Science 269 (1995) 5221.

\bibitem{Dav95}
K.\ B.\ Davis et. al., 
Phys.\ Rev.\ Lett.\ 75 (1995) 3969.

\bibitem{Gre02}
M.\ Greiner, et al., Nature 415 (2002) 39.

\bibitem{Tro10}
S.\ Trotzky et. al., Nature Phys.\ 6 (2010) 998.

\bibitem{Llo96}
S.\ Lloyd, Science 273 (1996) 1073.

\bibitem{Jak98}
D.\ Jaksch et. al.,
Phys.\ Rev.\ Lett.\ 81 (1998) 3108.

\bibitem{Cir12}
J.\ I.\ Cirac and P.\ Zoller, Nature Phys.\ 8 (2012) 264.

\bibitem{Lew12}
M.\ Lewenstein, A.\ Sanpera, and V.\ Ahufinger, ``Ultracold Atoms in Optical 
Lattices: Simulating Quantum Many-Body Systems'', Oxford University Press 
(2012).

\bibitem{Blo12} 
I.\ Bloch, J.\ Dalibard, and S.\ Nascimbene, Nature Phys.\ 8 (2012) 267.

\bibitem{Bla12}
R.\ Blatt and C.\ F.\ Ross, Nature Phys.\ 8 (2012) 277.

\bibitem{Asp12}
A.\ Aspuru-Guzik and P.\ Walther, Nature Phys.\ 8 (2012) 285.

\bibitem{Hou12}
A.\ A.\ Houck, H.\ E.\ T\"ureci, and J.\ Koch, Nature Phys.\ 8 (2012) 292.

\bibitem{Kap11}
E.\ Kapit and E.\ Mueller, Phys.\ Rev.\ A83 (2011) 033625.

\bibitem{Szi11}
G.\ Szirmai, E.\ Szirmai, A.\ Zamora, and M.\ Lewenstein,
Phys.\ Rev.\ A84 (2011) 011611.

\bibitem{Zoh12}
E.\ Zohar, J.\ Cirac, and B.\ Reznik, Phys.\ Rev.\ Lett.\ 109 (2012) 125302.

\bibitem{Ban12}
D.\ Banerjee et. al., Phys.\ Rev.\ Lett.\ 109 (2012) 175302.

\bibitem{Ban13}
D.\ Banerjee et. al., Phys.\ Rev.\ Lett.\ 110 (2013) 125303.

\bibitem{Zoh13}
E.\ Zohar, J.\ Cirac, and B.\ Reznik, Phys.\ Rev.\ Lett.\ 110 (2013) 125304.

\bibitem{Tag13a}
L.\ Tagliacozzo et. al., 
Nature Commun.\ 4 (2013) 2615.

\bibitem{Tag13b}
L.\ Tagliacozzo, A.\ Celi, A.\ Zamora, and M.\ Lewenstein,
Ann.\ Phys.\ 330 (2013) 160. 

\bibitem{Wie13}
U.-J.\ Wiese, Annalen der Physik 525 (2013) 777.

\bibitem{Gri84}
R.\ B.\ Griffiths, J.\ Stat.\ Phys.\ 36 (1984) 219.

\bibitem{Sch61}
J.\ Schwinger, J.\ Math.\ Phys.\ 2 (1961) 407.

\bibitem{Kel65}
L.\ V.\ Keldysh, JETP 47 (1965) 1515.

\bibitem{Kos72}
A.\ Kossakowski, Rep.\ Math.\ Phys.\ 3 (1972) 247.

\bibitem{Lin76}
G.\ Lindblad, Commun.\ Math.\ Phys.\ 48 (1976) 119.

\bibitem{Kra83}
K.\ Kraus, States, Effects and Operations, Fundamental Notions of Quantum 
Theory, Academic, Berlin (1983).

\bibitem{Ban14}
D.\ Banerjee, F.\ -J.\ Jiang, M.\ Kon, and U.\ -J.\ Wiese,
 Phys.\ Rev.\ B \ 90, 241104 (2014).

\bibitem{Flo15}
F.\ Hebenstreit et. al., Phys. Rev. B92, 035116 (2015).

\bibitem{Hoh77}
P.\ C.\ Hohenberg and B.\ I.\ Halperin, Rev.\ Mod.\ Phys.\ 49 (1977) 436.

\bibitem{San08}
A.\ W.\ Sandvik and H.\ G.\ Evertz, Phys.\ Rev.\ B82 (2010) 024407. 

\bibitem{Ger09}
U.\ Gerber et. al., J.\ Stat.\ Mech.\ (2009) P03021.

\bibitem{Flo15a}
D.\ Banerjee, F.\ Hebenstreit, F.\ -J.\ Jiang, and U.\ -J.\ Wiese,
Phys. Rev. B92, 121104 (2015).

\end{thebibliography}
\end{document}